**Long-Term Changes in the Variability of Pulsating Red Giants (and One R CrB Star)**

**John R. Percy**
Department of Astronomy and Astrophysics, and Dunlap Institute of Astronomy and Astrophysics, University of Toronto, Toronto, ON M5S 3H4, Canada; john.percy@utoronto.ca

**Arthur Lei Qiu**
Department of Astronomy and Astrophysics, University of Toronto, Toronto, ON M5S 3H4, Canada; arthur.qiu@mail.utoronto.ca

**Abstract.**  We have used many decades of visual observations from the AAVSO International Database, and the AAVSO time-series analysis package VSTAR, to study the long-term changes in period, amplitude, and mean magnitude in about 30 normal pulsating red giants (PRGs) i.e. those without large secular changes in period, as well as a few of the rare PRGs which *do* have such secular period changes.  The periods of the typical PRGs "wander" on time scales of about 40 pulsation periods – significantly longer than the time scales of amplitude variation which are 20-35, with a mean of 27 pulsation periods.  We have also studied the range and time scale of the long-term changes in pulsation amplitude and mean magnitude, as well as period, and looked for correlations between these.  Changes in mean magnitude are larger in stars with larger mean amplitude, but correlate negatively with *changes* in amplitude.  There is a weak positive correlation between the long-term period changes and amplitude changes.  The causes of these three kinds of long-term variations are still not clear.  We note, from the presence of harmonics in the Fourier spectra, that the longest-period PRGs have distinctly non-sinusoidal phase curves.  For studying PRGs, we demonstrate the advantage of studying stars with minimal seasonal gaps in the observations, such as those near the celestial poles.  We studied Z UMi, misclassified as a possible Mira star but actually an RCB star.  We determined times of onset of its fadings, but were not able to determine a coherent pulsation period for this star at maximum, with an amplitude greater than 0.05.  We did, however, find that the times of onset of fadings were "locked" to a 41.98-day period – a typical pulsation period for an R CrB star.

**1. Introduction**

When low- to medium-mass stars exhaust their hydrogen and helium nuclear fuel, they expand and cool, and become red giants or asymptotic-branch (AGB) stars.  In this paper, we shall lump these together as red giants.

Red giants are unstable to radial pulsation.  As they expand, their pulsation period increases from days to hundreds of days.  Their visual amplitude increases from hundredths of a magnitude, to up to 10 magnitudes.

In the *General Catalogue of Variable Stars* (GCVS; Samus *et al.* 2017), pulsating red giants (PRGs) are classified according to their light curves.  Mira (M) stars have reasonably regular light curves, with visual peak-to-peak amplitudes greater than 2.5 magnitudes.  Semi-regular (SR) stars are classified as SRa if there is appreciable periodicity, and SRb if there is very little periodicity.  Irregular (L) stars have no periodicity.   Percy and Kojar (2013), Percy and Long (2010), Percy and Tan (2013), and Percy and Terziev (2011) have published detailed analyses of AAVSO observations of SRa, SRb, and L stars.

There are several processes which can contribute to non-periodicity or apparent irregularity in PRG light curves, including the following:

- In some stars, both the fundamental and first overtone pulsation mode are excited (Kiss *et al* 1999).  The period ratios can be used to derive potentially useful astrophysical information (Percy and Huang 2015).

- The periods of PRGs "wander" by several percent on time scales of decades (Eddington and Plakidis 1929, Percy and Colivas 1999). This phenomenon can be modelled by random, cycle-to-cycle period fluctuations.
- About a third of all PRGs show long secondary periods (LSPs), 5-10 times the pulsation period depending on the pulsation mode (Wood 1999). The nature and cause of LSPs are unknown.
- The amplitudes of PRGs vary by up to a factor of 10 on time scales of 20-30 pulsation periods (Percy and Abachi 2013, Percy and Laing 2017).
- In a very few stars, thermal pulses cause large, secular changes in period, amplitude, and mean magnitude (Templeton *et al*. 2008 and references therein)

These processes occur on time scales which are much longer than the pulsation period, which itself can be hundreds of days. Since visual observations of these stars have been made for many decades, these observations – despite their limitations – are the best tool for studying long-term changes in the variability parameters of these stars. The purpose of this project was to use such visual observations to obtain further information about the long-term changes in period, amplitude, and mean magnitude in a sample of PRGs, and any correlations between these.

## 2. Data and analysis

We analyzed visual observations from the AAVSO International Database (AID: Kafka 2018) using the AAVSO's VSTAR software package (Benn 2013). It includes both a Fourier and wavelet analysis routine. The latter uses the Weighted Wavelet Z-Transform (WWZ) method (Foster 1996). The wavelet scans along the dataset, estimating the most likely value of the period and amplitude at each point in time, resulting in graphs which show the best-fit period and amplitude versus time.

## 3. Results and discussion

3.1 An alternate way of quantifying the "wandering" periods of pulsating red giants

The wandering periods in PRGs have been known for over a century, and can be modelled as random cycle-to-cycle period fluctuations (Eddington and Plakidis 1929). This implies a process which takes place on a time scale of approximately one pulsation period. One could also look at the period-versus-time graphs in a more global way, assuming them to represent long-term changes, and then to measure the typical time scale of the variations, in a similar way as was done for the amplitude-versus-time graphs (Percy and Abachi 2013, and especially Percy and Laing 2017). We used the first 20 (O-C) diagrams of Karlsson (2013), rather than period-versus-time plots, to measure the ratio of L, the length of the cycles of period increase and decrease, to the pulsation period P. This same analysis could have been done with wavelet analysis; both it and the (O-C) method can be used to display and measure cycles of period increase and decrease. Conveniently, the Karlsson (O-C) diagrams measure time in units of pulsation periods. See Percy and Abachi (2013) and Percy and Laing (2017) for a discussion of the uncertainties of determining L. The values of L/P were as follows: R And (35), T And (51), UU And (30), V And (38), W And (37), X And (32), Y And (67), RR And (60), RW And (44), SV And (35), SX And (35), SZ And (49), TU And (56), UZ And (40), V Ant (56), T Aps (40), R Aqr (28), S Aqr (133), T Aqr (60), W Aqr (40). The median value of L/P is about 40. This ratio is significantly larger than that for amplitude increases and decreases (20-35, mean 27), and much larger than the ratio of LSP/P (5-10) i.e. the time scales are different. We emphasize, though, that the wandering periods may still be a result of accumulated cycle-to-cycle fluctuations.

3.2 Measuring the changing mean magnitudes of pulsating red giants

We have previously studied the long-term changes in the periods and amplitudes of PRGs, but not the mean magnitudes. Some stars have LSPs, of course, but we wondered whether there were even longer-term variations in mean magnitude, an order of magnitude longer, possibly correlated with long-term variations in period or amplitude. Both the light curves and the Fourier spectra suggest that such variations might be present.

One complication is the possible interaction of the pulsational variations and the seasonal gaps. It can produce apparent long-term variations in mean magnitude. These correspond to alias peaks in the Fourier spectrum which lie close to zero frequency. One strategy would be to analyze stars with minimal seasonal gaps, those near the celestial poles; we have done this in section 3.4.

We used the wavelet routine within VSTAR to determine and graph the long-term changes in mean magnitude. Although mean magnitude is not directly gra phed in VSTAR, the necessary data can be extracted from the tables produced by VSTAR. The results of these are contained in Table 1, and examples are shown in Figure 1. Graphs like these were constructed for all the stars in Table 1, and used to determine the changes and ranges in the period P, the peak-to-peak amplitude A, and the mean magnitude M. The graphs were also used to assess the correlation between the variations.

The values of ΔM cluster between 0.3-0.6 and 0.7-0.9. It is not clear whether the bimodal distribution is significant.

The time scales for the long-term changes in mean magnitude, when quantified in the same way as for the changes in period (section 3.1) and amplitude, give time scales in the range of 20-30 pulsation periods. Since this is also the time scale of amplitude variation, this raises the concern that the mean magnitude variations might be artifacts of the amplitude variations. We also note that ΔM correlates with the mean amplitude A. This may be because both are correlated with some more fundamental parameter, such as temperature. In PRGs, period and amplitude generally increase as temperature decreases. Large ΔM stars are all long-period stars; smaller ΔM stars occur at all periods.

3.3 Correlations between changing period, amplitude, and mean magnitude?

We compared the long-term changes in period, amplitude, and mean magnitude, and qualitatively assessed, by eye, whether there appeared to be a positive correlation, a negative correlation, or no correlation at all i.e whether, with time, they tended to change in the same direction. The changes in period and amplitude were determined using wavelet analysis, and are expressed as the total range in P (ΔP), A (ΔA), and M (ΔM). ΔP increases with P, as might be expected; ΔA increases slightly with P; longer-period stars tend to have larger amplitudes. ΔM does not increase or decrease with P, but lies in the range 0.2 to 0.8. These results are represented by the symbols +, -, and 0, respectively, in Table 1.

There is a very weak positive correlation between ΔP and ΔA, and a weak negative correlation between ΔA and ΔM. This is discussed further below.

3.4 The advantages of Ursa Minor and Octans

Visual observations such as those in the AID normally contain seasonal gaps, because the star is unavailable for viewing at certain times of the year, depending on its position in the sky. These seasonal gaps produce alias peaks in the Fourier spectrum, due to the one-year periodicity of the times of observations. The alias peaks are frequencies of f ± N/365.25 where f is the true frequency. The strongest alias peaks are at N = 1. See Percy (2015) for a discussion. For pulsating red giants, the alias peaks can be confused with harmonic or overtone periods.

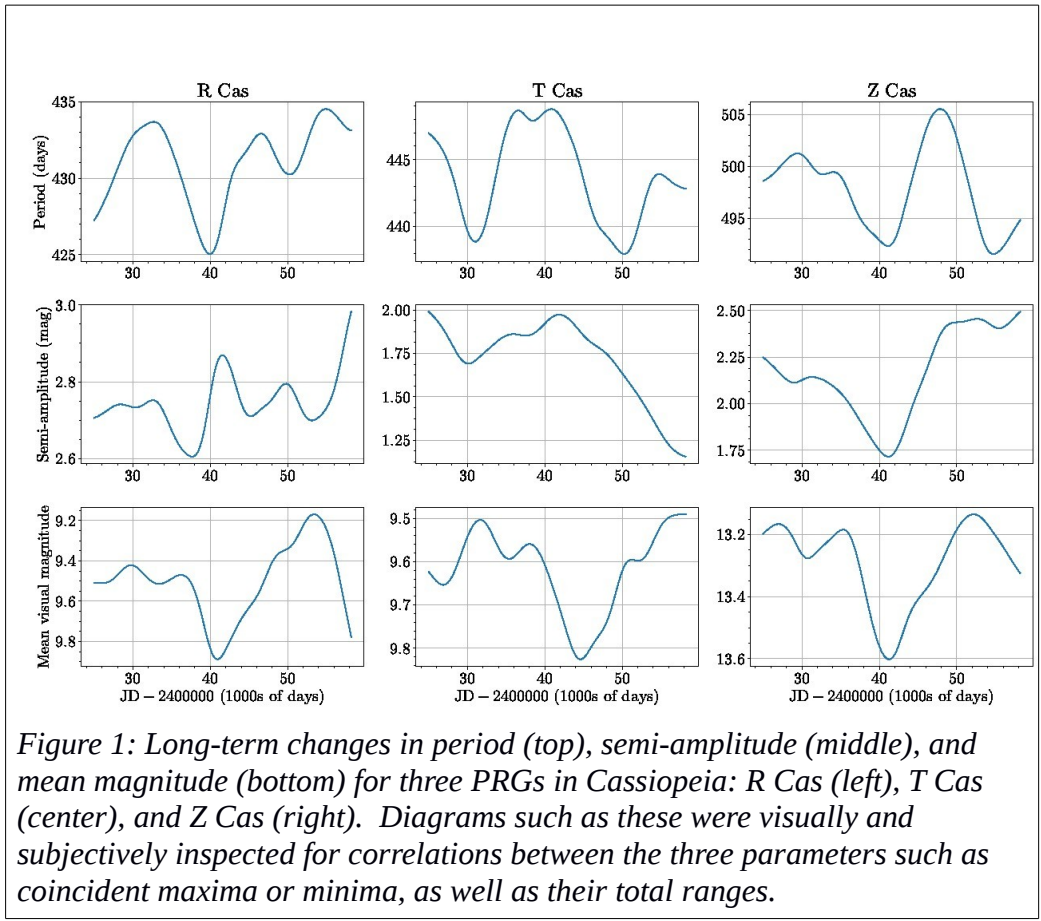

*Figure 1: Long-term changes in period (top), semi-amplitude (middle), and mean magnitude (bottom) for three PRGs in Cassiopeia: R Cas (left), T Cas (center), and Z Cas (right). Diagrams such as these were visually and subjectively inspected for correlations between the three parameters such as coincident maxima or minima, as well as their total ranges.*

Stars near the celestial poles tend to have minimal seasonal gaps, because the stars can be observed all year, without interference from the sun. Ursa Minor and Octans are constellations near the north and south celestial poles, respectively. Figures 2 and 3 compare the DCDFT spectra for U Cnc and R Oct. The former is near the ecliptic, and shows a complex spectrum with the pulsation frequency 0.003278 cycles per day (cpd) or period 306 days, alias frequencies at 0.000542 and 0.006014 cpd, harmonic frequencies at 0.006554 and 0.009794 cpd, and aliases of the first harmonic at 0.003818 and 0.009296 cpd, and a possible first overtone at 0.007088 cpd. S Oct is near the south celestial pole, and shows only the pulsation frequency 0.002468 cpd (period 405.2 days) and harmonic frequencies at two, three, and four times this.

As noted earlier, the study of these stars was motivated by the concern that interaction between the pulsational variations and the seasonal gaps might produce apparent low-frequency variability in mean magnitude. In fact, the long-term variability (ΔM) of the seven stars in Oct and UMi is similar to that of the other stars, in both total range and time scale. Whether this variability is real, or due to the distribution of the observations over the pulsation cycle or some other observational factor, or a combination of the two, we cannot tell.

As for correlations: there is a tendency for ΔP and ΔA to be positively correlated, ΔA and ΔM to be negatively correlated, and ΔP and ΔM to be uncorrelated. There were similar but weaker correlations among the stars not in Oct or UMi. These correlations are suggestive, but are not present in every star.

*Table 1: Long-term changes in the period, amplitude, and mean magnitude of pulsating red giants, and correlations between these. P = period; A = amplitude, and M = mean magnitude, and ΔP, ΔA, and ΔM are the total ranges in period, amplitude, and mean magnitude..*

| Star | P(days) | A | ΔP(days) | ΔA | ΔM | σPA | σPM | σAM |
| --- | --- | --- | --- | --- | --- | --- | --- | --- |
| R And | 410 | 3.05 | 12.90 | 0.38 | 1.34 | 0 | 0 | 0 |
| T And | 281 | 2.07 | 9.46 | 0.35 | 0.36 | + | 0 | 0 |
| V And | 256 | 2.12 | 7.04 | 0.47 | 0.44 | + | - | - |
| X And | 343 | 2.52 | 8.50 | 0.76 | 0.33 | 0 | 0 | 0 |
| RR And | 331 | 2.88 | 4.46 | 0.49 | 0.42 | 0 | 0 | 0 |
| RW And | 430 | 2.92 | 9.65 | 0.80 | 0.72 | 0 | 0 | - |
| SV And | 313 | 1.97 | 9.18 | 0.59 | 0.53 | 0 | 0 | - |
| TU And | 313 | 1.77 | 10.24 | 0.45 | 0.56 | - | - | 0 |
| UW And | 237 | 1.83 | 5.60 | 0.53 | 0.30 | + | 0 | 0 |
| YZ And | 207 | 2.15 | 4.22 | 0.51 | 0.64 | 0 | 0 | 0 |
| R Aqr | 386 | 1.74 | 12.29 | 1.76 | 1.03 | + | - | - |
| S Boo | 270 | 1.77 | 8.60 | 0.35 | 0.30 | 0 | 0 | - |
| R Car | 310 | 2.33 | 5.56 | 0.36 | 0.23 | 0 | 0 | 0 |
| S Car | 151 | 1.19 | 2.99 | 0.45 | 0.47 | + | - | - |
| R Cas | 430 | 2.60 | 9.47 | 0.36 | 0.72 | - | 0 | + |
| S Cas | 608 | 2.26 | 15.50 | 0.89 | 0.94 | - | - | + |
| T Cas | 445 | 1.54 | 10.82 | 0.83 | 0.34 | 0 | + | - |
| U Cas | 277 | 2.71 | 5.44 | 0.55 | 0.33 | + | 0 | - |
| Z Cas | 497 | 2.03 | 13.90 | 0.76 | 0.47 | + | + | + |
| TY Cas | 645 | 2.11 | 15.15 | 0.89 | 0.58 | - | + | - |
| R Cen | 502 | 0.81 | 50.40 | 1.14 | 0.20 | + | 0 | 0 |
| R Oct | 405 | 1.86 | 13.60 | 0.73 | 0.19 | (+) | 0 | 0 |
| S Oct | 259 | 2.55 | 3.93 | 0.83 | 1.04 | (+) | 0 | 0 |
| T Oct | 219 | 1.65 | 5.98 | 0.91 | 1.00 | 0 | 0 | 0 |
| U Oct | 303 | 2.44 | 8.10 | 0.27 | 0.77 | (+) | 0 | - |
| R UMi | 324 | 0.43 | 13.80 | 0.45 | 0.31 | 0 | + | - |
| S UMi | 327 | 1.31 | 13.10 | 0.47 | 0.63 | + | 0 | - |
| U UMi | 325 | 1.25 | 12.00 | 0.42 | 0.27 | + | 0 | 0 |

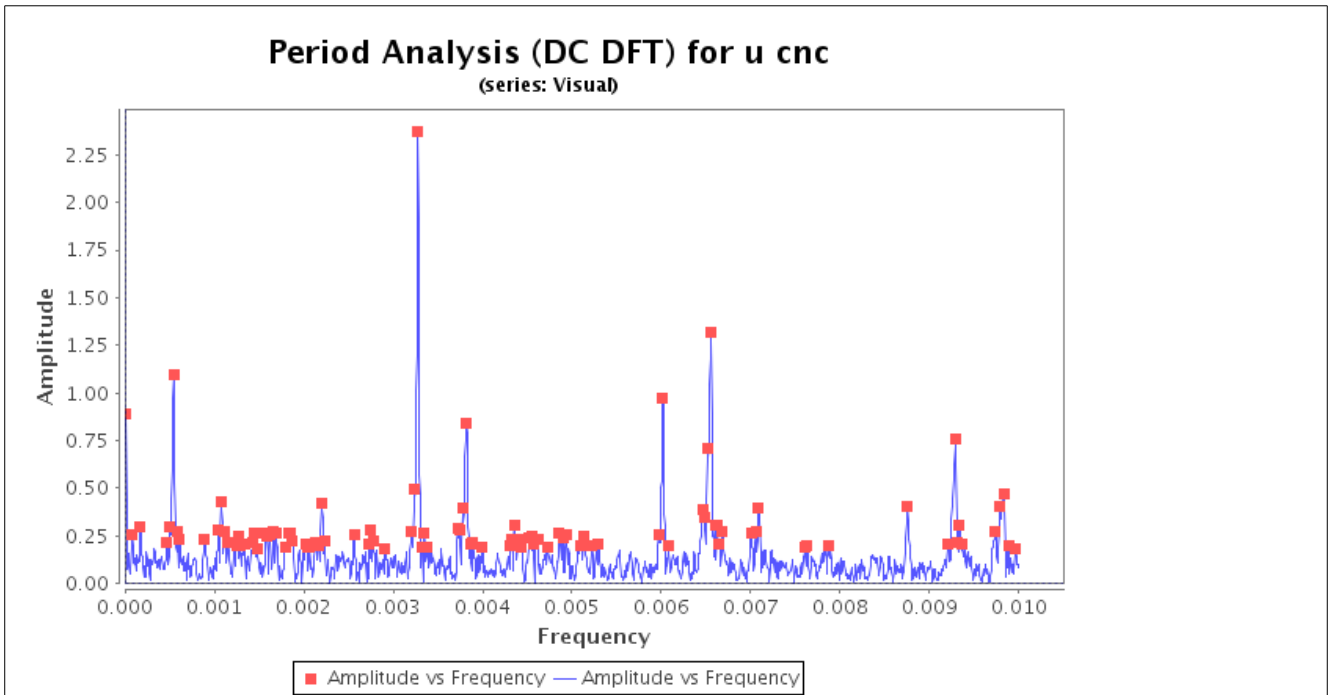

*Figure 2: The Fourier spectrum of U Cnc, a star near the ecliptic with significant seasonal gaps, and therefore alias peaks in the spectrum. See text for identification of the alias, harmonic, and overtone peaks.*

3.5 Stars with significant secular period change

The vast majority of PRGs have wandering periods, but a few percent have periods that change secularly and significantly (Templeton *et al*. 2008), probably due to a thermal pulse. Because they are so unusual, these stars have previously been studied in some detail; see discussion and references in Templeton *et al*. (2008). For completeness, we list nine of these, in Table 2. The correlations between changes in P, A, and M, as given in the last three columns, are qualitative and based on visual inspection.
There is a generally positive correlation between amplitude and period change, and a negative correlation between mean magnitude change, and period or amplitude change.

3.6 The nature of Z UMi

In the course of undertaking the study of the stars in UMi and Oct, we came upon Z UMi. It is classified in the *General Catalogue of Variable Stars* (Samus *et al.* 2017) as M: i.e. a possible Mira star, with a period of 475 days. At first inspection, the light curve (Figure 5) – especially the early part -- bears some resemblance to a Mira star but, on second inspection, is clearly that of an R Coronae Borealis star. Indeed, it was identified as a new R CrB star by Benson *et al*. (1994). Fourier analysis gives strongest peaks at "periods" of 1351 and 895 days, but these are just the best fits to the random fadings; they have no physical significance.

*Table 2: Variability properties, their long-term changes, and directions, and correlations between these, for PRGs with significant secular period changes.*

| Star | P(days) | A | ΔP(day) | ΔA | ΔM | σPA | σPM | σAM |
|---|---|---|---|---|---|---|---|---|
| R Aql | 282.6 | 0.83 | 55↓ | 0.8↓ | 0.7↑ | + | - | - |
| R Cen | 546.1 | 0.81 | 50↓ | 1.1↓ | 0.0 | + | (-) | (-) |
| BF Cep | 429.3 | 1.74 | 14↑ | 0.4↑ | 0.1↓ | + | - | - |
| BH Cru | 520.6 | 1.18 | 35↑ | 0.5↑ | 0.3↓ | + | - | - |
| LX Cyg | 565.3 | 1.05 | 100↑ | - | 0.8↓ | 0 | 0 | 0 |
| W Dra | 279.8 | 0.93 | 33↑ | 0.8↑ | 1.0↓ | 0 | - | 0 |
| R Hya | 388.0 | 1.08 | 50↓ | 0.8↓ | 0.6↑ | + | (-) | - |
| Z Tau | 460.0 | 0.59 | 40↓ | - | 1.0↑ | 0 | - | 0 |
| T UMi | 312.2 | 0.56 | 120↓ | 2.0↓ | 1.0↑ | + | - | - |

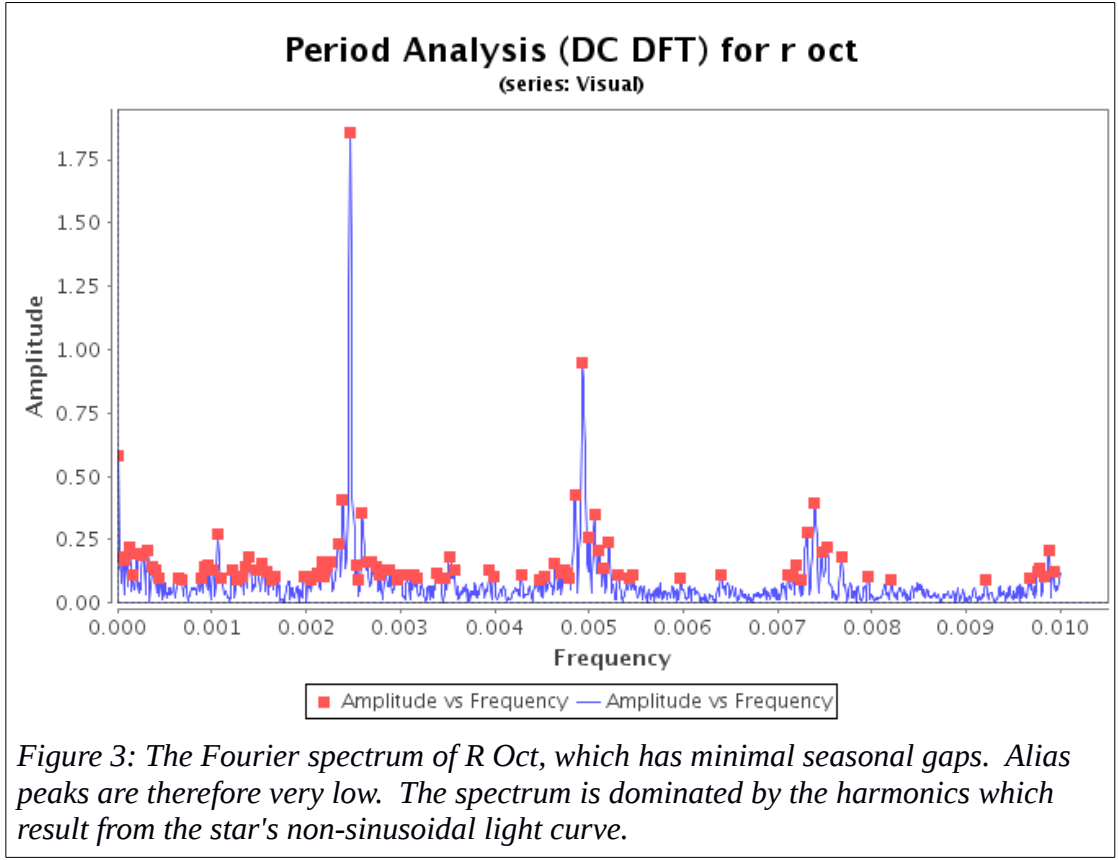

*Figure 3: The Fourier spectrum of R Oct, which has minimal seasonal gaps. Alias peaks are therefore very low. The spectrum is dominated by the harmonics which result from the star's non-sinusoidal light curve.*

Many R CrB stars display low-amplitude pulsations at maximum; see Table 3 in Rao and Lambert (2015) for a list. They give a pulsation period of 130 days for Z UMi, based on very limited photoelectric measurements by Benson *et al.* (1994). We therefore examined AAVSO visual and V observations in the few intervals when the star was near maximum light. No periods with amplitudes greater than 0.05 stand out, though there are some suggestions of low-amplitude variations on time scales of 50-100 days (figure 6).

We also compiled a list of times of onset of fadings (Table 3), to see if they were "locked" to some period which might be a pulsation period, as has been found in at least five R CrB stars (Crause *et al.* 2007 and references therein). Such a "lock" might imply a causative relation between pulsation and the onset of fading. Indeed, they appear to be locked to a period of 41.98 days. Table 3 lists the times of onset, the cycle numbers of the 41.98-day period, and the values of (O-C) expressed in periods. The average absolute value of the (O-C) is 0.09 cycles, or 4 days. This is similar to a value obtained by Crause *et al.* (2007) in five other R CrB stars whose fadings were locked to a pulsation period.

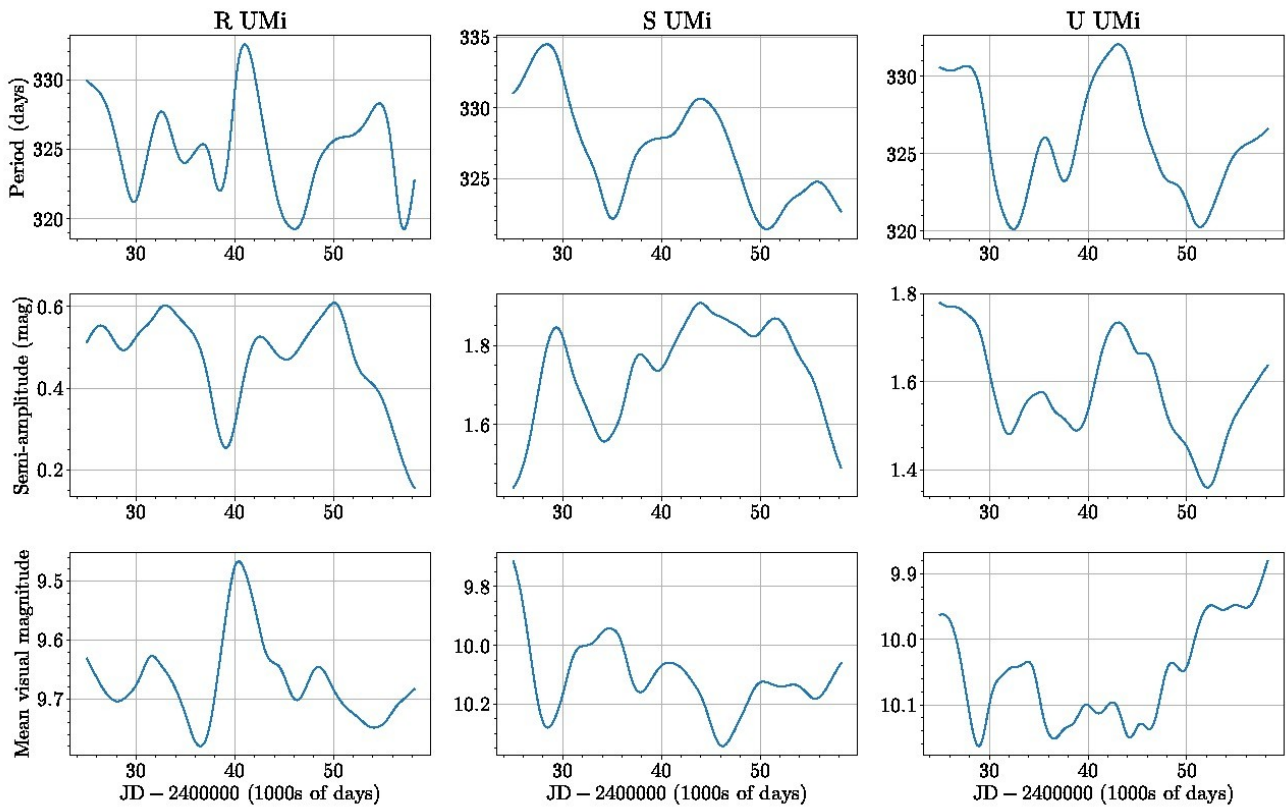

*Figure 4: Long-term changes in period (top), semi-amplitude (middle), and mean magnitude (bottom) for three stars in Ursa Minor: R UMi (left), S UMi (center), and U UMi (right). These stars have minimal seasonal gaps to affect the analysis.*

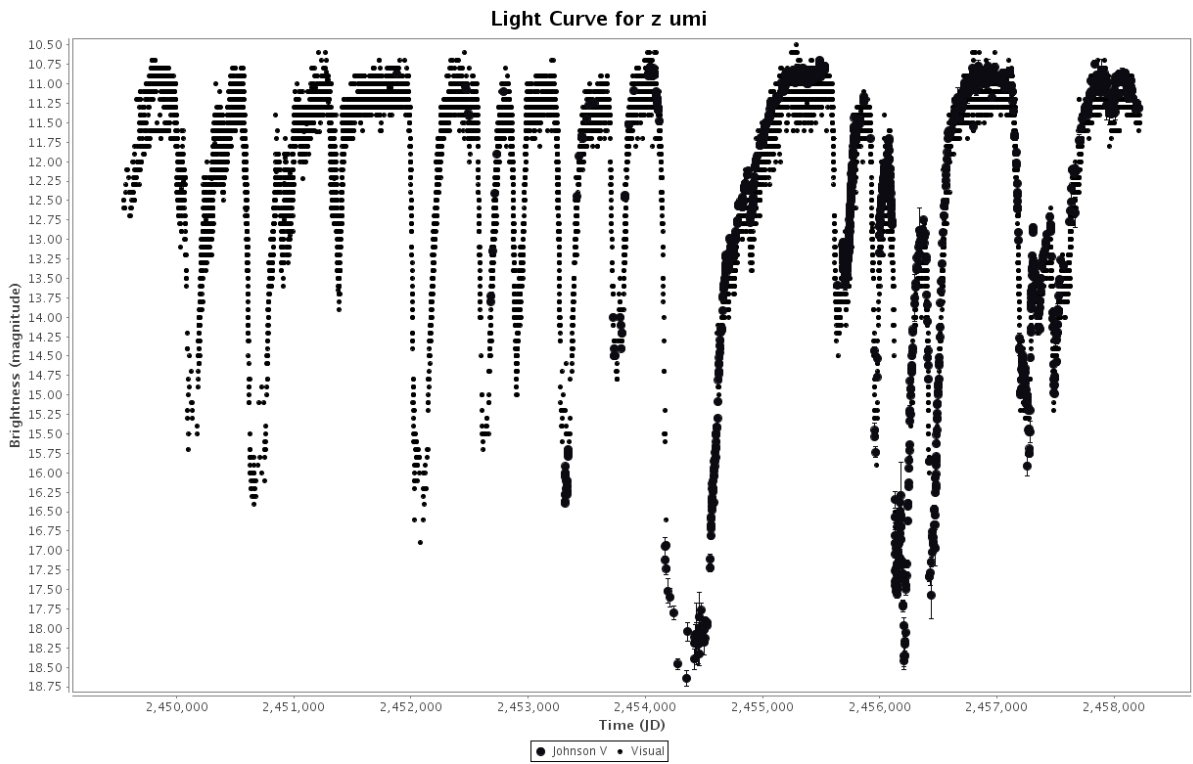

*Figure 5: Light curve of Z UMi, from visual observations in the AAVSO International Database. The early variability, especially if sparsely-sampled, bears some resemblance to that of a Mira star.*

3.7 Phase curves of pulsating red giants

Most PRGs have reasonably sinusoidal phase curves. We note, however, that PRGs with longer periods tend to have non-sinusoidal phase curves. This includes : RW Lyr (503), Z Pup (510), V Cam (523), V Del (528), and S Cas (613), TY Cas (645); the numbers in brackets are the pulsation periods in days. As a way to quantify the non-sinusoidal nature, we used DCDFT in VSTAR to determine the ratio of the first-harmonic amplitude to the fundamental amplitude. The ratio ranges from 0.45 in RW Lyr, to 0.78 in V Del. RW Lyr, incidentally, varies in amplitude by a factor of two – unusually large for a Mira star.

We note also that the phase curves of larger-amplitude PRG LSPs are often non-sinusoidal; see figures 1 and 4 in Percy and Deibert (2016), for instance, which show the LSP light curves of U Del and Y Lyn. The similarity of the phase curves may be entirely coincidal; there is no evidence otherwise.

**4. Discussion**

It is not clear whether the long-term variations in mean magnitude of the PRGs are spurious, or real, and, if they are real, what the cause is. If spurious, they could arise from the random distribution of the observations over the various variatility cycles, or changes in the visual observers and their characteristics over time, or due to changes in the calibration of the visual photometry system (though the AAVSO tries very hard to avoid such changes). If real, they could reflect some long-term variation in the physical properties of the star, perhaps due to the convection process, or variations in the amount of obscuring dust around the star. It would be helpful to do a cycle-by-cycle analysis, perhaps using stars with minimal seasonal gaps.

The causes of the LSPs, the longer-term variations in period, and in amplitude are also not known; see discussion in Percy and Deibert (2016). The variations in period have traditionally been ascribed to random cycle-to-cycle fluctuations, but there is no physical evidence for this. Giant convection cells may be involved in these variations, either through their turnover, or through rotational variability. Large granulation cells have recently been imaged on the surface of π¹ Gruis, a PRG (Paladini *et al.* 2018).

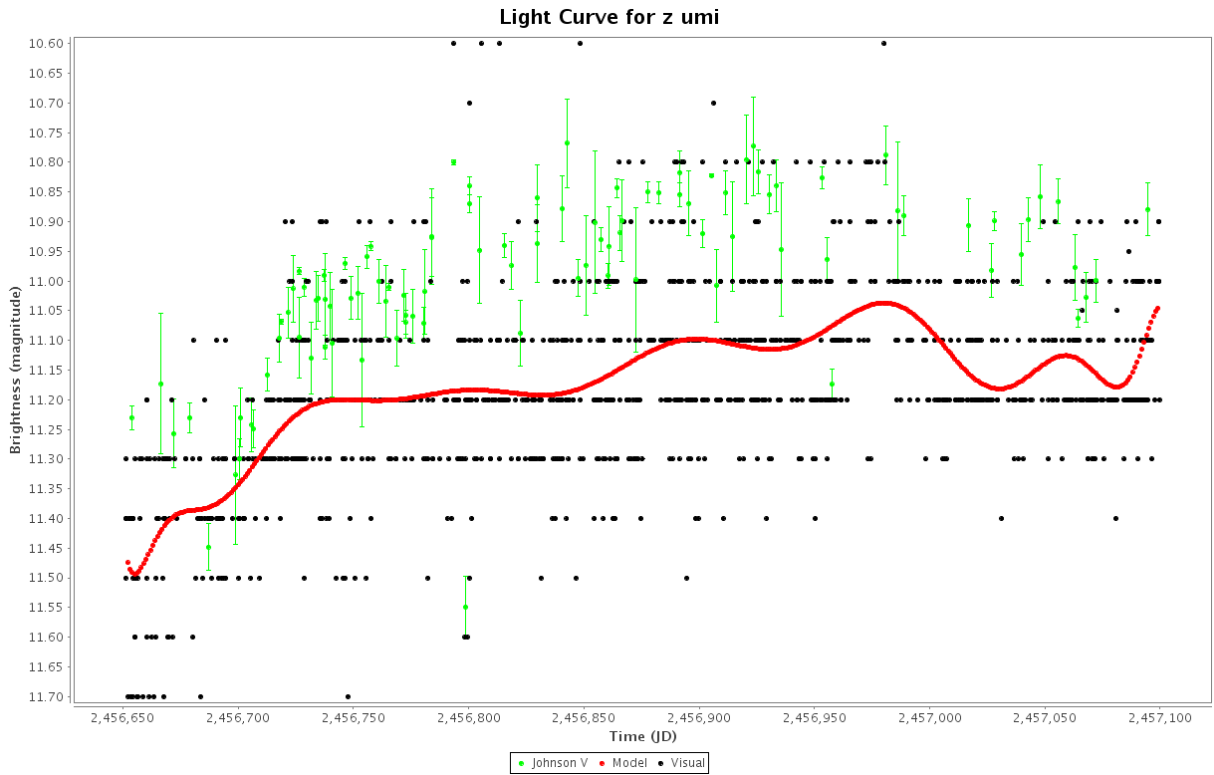

*Figure 6: The light curve of Z UMi near maximum. The small black points are visual. The solid red line is a polynomial fit to these. The green points with error bars are Johnson V observations. Note the possible low-amplitude variability with a time scale of a few tens of days. The variability is marginal, and there is no evidence for or against periodicity.*

*Table 3: Times of onset of fadings in Z UMi, and (O-C) analysis of these for a period of 41.98 days and an initial epoch of JD 2450004.*

| Onset JD | Cycle | (O-C)/P |
|---|---|---|
| 2450004 | 0 | 0.00 |
| 2450557 | 13 | 0.17 |
| 2451300 | 31 | -0.13 |
| 1451977 | 47 | 0.00 |
| 2452572 | 61 | 0.17 |
| 2452861 | 68 | 0.05 |
| 2453239 | 77 | 0.06 |
| 2453699 | 88 | 0.02 |
| 2454112 | 98 | -0.14 |
| 2455590 | 133 | 0.06 |
| 2455921 | 141 | -0.05 |
| 2456086 | 145 | -0.12 |
| 2456394 | 152 | 0.22 |
| 2457139 | 170 | -0.04 |
| 2457937 | 189 | -0.03 |
| 2458103 | 193 | -0.07 |

**Conclusions**

We present some new analyses of PRGs, including some interesting *possible* correlations between the long-term variations in the periods, amplitudes, and mean magnitudes. Given the complexity of these stars' variations, and the limitations of visual data, we cannot say more. A much larger study, possibly with a more quantitative comparison between the variations, might possibly confirm these correlations

**6. Acknowledgements**

We thank the AAVSO observers who made the observations on which this project is based, the AAVSO staff who archived them and made them publicly available, and the developers of the VSTAR package which we used for analysis. This paper is based, in part, on a research project carried out by undergraduate math and science student ALQ. We acknowledge and thank the University of Toronto Work-Study Program for existing, and for financial support. This project made use of the SIMBAD database, maintained in Strasbourg, France. The Dunlap Institute is funded through an endowment established by the David Dunlap family and the University of Toronto.